\documentclass[12pt]{article}
\begin{document}\def\p{\phi}\def\P{\Phi}\def\a{\alpha}\def\e{\varepsilon}
\def\be{\begin{equation}}\def\ee{\end{equation}}\def\l{\label}\def
\0{\setcounter{equation}{0}}\def\b{\beta}\def\S{\Sigma}\def\C{\cite}
\def\r{\ref}\def\ba{\begin{eqnarray}}\def\ea{\end{eqnarray}}\def\n{\nonumber
}
\def\R{\rho}\def\X{\Xi}\def\x{\xi}\def\la{\lambda}\def\d{\delta}\def\s{\sigm
a}
\def\f{\frac}\def\D{\Delta}\def\pa{\partial}\def\Th{\Theta}\def\o{\omega}
\def\O{\Omega}\def\th{\theta}\def\ga{\gamma}\def\Ga{\Gamma}\def\t{\times}
\def\h{\hat}\def\rar{\rightarrow}\def\vp{\varphi}\def\inf{\infty}\def\le{\le
ft}
\def\ri{\right}\def\foot{\footnote}

\title{Topology and perturbation theory}
\author{ J.Manjavidze$^{a)}$}

\maketitle

\begin{abstract}
Paper contains description of the fields nonlinear modes successive
quantization scheme. It is shown that the path integrals for
absorption part of amplitudes are defined on the Dirac ($\d$-like)
functional measure.  This permits arbitrary transformation of the
functional integral variables. New form of the perturbation theory
achieved by mapping the quantum dynamics in the space $W_G$ of the
({\it action, angle})-type collective variables.  It is shown that
the transformed perturbation theory contributions are accumulated
exactly on the boundary $\pa W_G$. Abilities of the developed
formalism are illustrated by the Coulomb problem. This model is
solved in the $W_C$=({\it angle, angular momentum, Runge-Lentz
vector}) space and the reason of its exact integrability is
`emptiness' of $\pa W_C$.
\end{abstract}


\section{Introduction}\0

Solution of great number of the modern field-theoretical problems
rest on absence of workable perturbation theory in the vicinity
of actions nontrivial extremum $u_c(x)$. The think is that the
dynamics of perturbations in such fields is rather complicate $^1$.
So, for instance, beyond the semiclassical approximation of path
integrals one should know solution of the equation:
\be
(\pa^2 + v''(u_c))_x G(x,x';u_c)=\d(x-x').
\l{i3}\ee
for the Green function $G(x,x';u_c)$. The exact solution of this
equation is unknown since the operator $(\pa^2 + v''(u_c))_x$ is not
translationally invariant if $u_c =u_c(x)$. Of course, one can find
$G(x,x';u_c)$ perturbatively neglecting in the first approximation
the coordinate dependence in $v''(u_c)$. But this approximation is
applicable at small distances $|x-x'|\rar0$ only and the number of
modern speculations on the way as this restriction may be avoided is
enormous.  The perturbative QCD is an example of such solution.

A suspicion that the fields are not always useful variables arise
and an idea that the quantum theory formulated in other terms may be
much more effective seems natural (actually hoping that the
substitution may considerably simplify calculation of the integral).
I wish to show the quantitative realization of this idea and will
construct corresponding perturbation theory.  The main formal problem
on this way $^{2}$ consist in demonstration that the transformation
to new variables is unitary, i.e. conserves the $total$ probability.

So, the main goal of this paper is to formulate the perturbation
theory formalism for the case of nontrivial $u_c(x)$. Our
perturbation theory is nothing new if the field $u_c=const.$, but is
extremely effective for nontrivial $u_c(x)$. Actually the successive
approach to the strong-coupling perturbation theory is offered.

Having in mind the non-perturbative effects (field topological
excitations) the lattice decompositions are widely used. For instance,
number of problems of quantum mechanics was solved using the `time
sliced' method $^{3,4}$.  This approach presents the path integral
as a finite product of well defined ordinary Cauchy integrals and,
therefore, allows perform arbitrary transformations. But transformed
`effective' Lagrangian gains additional term $\sim\hbar^2$ in the
continuum limit.  Last one crucially depends from the way as the
`slicing' was performed and a general solution of this problem is
cumbersome. Our approach will not contain any ambiguities.

We will formulate the approach, risking to loose generality,
considering the simplest quantum-mechanical examples of particle
motion in the potential hole $v(u)$ with one non-degenerate minimum
at $u=0$. We will calculate the $probability$ $\R=\R (E)$ to find
the bound state with energy $E$. Namely, at the end we would solve
the plane Coulomb problem using our method.

Our experience may be useful for quantization of nonlinear waves
also. Indeed, introducing the convenient variables (collective
coordinates) one can reduce the quantum soliton-like excitations
problem to quantum-mechanical one.  This idea was considered
previously by many authors $^{5,6}$.

The aim of this paper is to show

(i)  {\it Origin of desired perturbation theory.}

The mechanism of unitary, i.e. the total probability conserved,
mapping $(u,p)(t)\rar(\x,\eta)(t)$, where $p(t)$ is the conjugate to
$u(t)$ momentum, of the functional measure on the space $W_G$ with
local coordinates $(\x,\eta)(t)$ is shown. It would be considered as
the factor space $W_G=G/\bar{G}$, where $G$ is the theory symmetry
group and its subgroup $\bar{G}$ is the symmetry of classical fields
$u_c=u_c(\x,\eta)$.

It is well known that if $J_i=J_i(u,p)$, $i=1,2,...,N$ are the first
integrals in involution then the $canonical$ transformation
$(u,p)\rar(J,Q)$ solves the mechanical problem (Liouville-Arnold
theorem). The $(u,p)_c$ flow is defined by the $2N$ system of coupled
algebraic equations
\be
\eta=J(u,p),~\x=Q(u,p).
\l{i14'}\ee

The mapping (\r{i14'}):
\be
J:T \rar W_G,
\l{i14}\ee
where $T$ is the $2N$-dimensional phase space and $W_G$ is a linear
space, introduces integral $manifold$ $J_{\o}=J^{-1} (\o)$ in such a
way that the $classical$ phase space flaw belongs to $J_{\o}$
$completely$.

Our methodological idea assumes quantization of the $J_{\o}$ manifold
instead of flow in $T$. This becomes possible iff the quantum
trajectory completely belongs to $J_{\o}$.  Last one means that
eqs.(\r{i14'}) have unique solution $(u,p)_c$ and $(\x,\eta)$ compose
a manifold.

The `direct' mapping (\r{i14}) assumes that $J$ is known.  But this
approach to general quantum problems seems inconvenient having in
mind the nonlinear modes quantization, when the number of degrees of
freedom $N=\infty$, or if the transformation is not canonical, see
(\r{i16}).  We will consider by this reason the `inverse' approach
starting from assumption that the classical flow exist.  Then we
would be able to reconstruct the motion in $W_G$ since $(u,p)_c$
belongs to $W_G$ completely.

In other words, we would like to describe the quantum dynamics in the
space of classical fields (orbits) $u_c$ parameters $(\x,\eta)(t)$.
The unitarity of such mapping is guaranteed be the fact that the
functional measure, as the consequence of the unitarity condition, is
Diracian $^{7}$:
\be
DM(u,p)=\d(E-H_T(u,p))\prod_t du(t)dp(t)\d (\dot{u} +
\frac{\pa H_j}{\pa p}) \d (\dot{p} - \frac{\pa H_j}{\pa u}),
\l{i1}\ee
where the Hamiltonian $$ H_j =\frac{1}{2}p^2 + v(u) -ju$$ includes
the energy of quantum fluctuations $ju$, with the provoking quantum
excitations force $j=j(t)$. Then the dynamical equilibrium between
ordinary mechanical forces ($\dot{p}(t)$, $-v'(u)$) and quantum force
($j(t)$) determined by $\d$-like measure (\r{i1}) allows to perform
arbitrary transformation of quantum measure, i.e. of $j$, caused by
transformations of $u$ and $p$.

A theory on such measure is `simple' since the functional
$\d$-function defines the complete set of contributions. So, the
constructive definition of the vacuum will be offered in Sec.IIC. It
will be shown the general method of the mapping, applicable for
field theories also, when $u_c=u_c(\vec{x};\x,\eta)$.

$Note$, there are in modern physics the remarkable attempts
to construct a geometrical approach to quantum mechanics
$^{8,9}$ and field theory $^{10}$. Our approach has an evident
geometrical interpretation and it will be widely used. It
has deal with the excitations of classical phase-space flows.  By
this reason, in contrast with above mentioned approaches, the
finite-dimensional manifolds only, as in classical mechanics, would
arise even in the field theories.

We would construct the mapping using the base of symplectic geometry.
Starting from assumption that $(\x,\eta)$ form the symplectic space
of arbitrary dimension we would demonstrate its projection on $W_G$.

Describing the perturbations of new dynamical variables
$(\x,\eta)(t)$ we take into account the quantum excitations of field
$u_c=u_c(\x,\eta)$. In considered below Coulomb problem the set
$(\x,\eta)=$(angle, angular momentum, length of Runge-Lentz vector)
unambiguously defines the Kepler orbits. Hence, the mapping
$(u,p)\rar(\x,\eta)$ is rightful since the quantum trajectory covers
$W_G=(\x,\eta)$ densely (fluctuations of $j(t)$ are defined on
Gauss measure) and since $(u,p)_c$ belong to $W_G$ completely.

(ii) {\it Structure of perturbation theory in the $G/\bar{G}$
space and as it can be applied.}

It will be shown that the quantum corrections of the transformed theory
are accumulated on the boundaries (bifurcation manifolds $^{11}$)
$\pa W_G$ of the factor space, i.e.  are defined mainly by $W_G$
$topology$. The important quantitative consequence would be the
observation that the quantum corrections may disappeared (totally or
partly) on $\pa W_G$ if the boundary is empty.

So, the problem of quantum corrections we reduce up to definition of
intersection of the boundary set $\{\pa u_c(\x,\eta)\}$ with the
boundary $\pa W_G$.  This circumstance would be useful for estimation
of quantum corrections. For all that the explicit form of $u_c$ is
not necessary since $\{\pa u_c\} \bigcap \pa W_G$ is estimated.

One should assume that $j(t)$ switched on adiabatically (in this case
we expand contributions in the vicinity of $j=0$) for effective use
of this definition of measure.  Otherwise we should know $j(t)$
exactly, including it into Lagrangian as the external field. The
measure would remain $\d$-like in last case also. So, the measure
(\r{i1}) allows to conclude that the solutions of classical equation

\be
\f{\d S(u)}{\d u(t)}=0.
\l{i2}\ee
defines the complete set of contributions $^{12}$.

Eq.(\r{i2}) reflects the ordinary Hamiltonian variational principle.
But the measure (\r{i1}) contains following additional information:

{\bf i.} Only $strict$ solutions of eq.(\r{i2}) should be taken into
account.

{\bf ii.} $\R(E)$ is described by the $sum$ of all solutions of
eq.(\r{i2}), independently from the value of corresponding
fluctuations;

{\bf iii.} $\R(E)$ did not contain the interference terms from various
topologically nonequivalent contributions. This displays the
orthogonality of corresponding Hilbert spaces;

{\bf iv.} The measure (\r{i1}) includes $j(t)$ as the `external
source';

{\bf v.} In the frame of above adiabaticity condition the field
$u(t)$ disturbed by $j(t)$ belongs to the same manifold (topology
class) as the classical field defined by (\r{i2}) $^{12}$.

One must underline that the measure (\r{i1}) is derived for
$real-time$ processes only, i.e.  is not valid for tunneling ones. By
this reason above conclusions should be taken carefully. The
corresponding selection rule will be given below in Sec.IIC.

The main results of this paper looks as follows.

({\bf A}) If the amplitude has the path integral representation
(\r{3.1}), then the unitarity condition leads to following
representation for $\R(E)$
\be
\R(E)=\int^\infty_0 dT e^{-i\h{K}(ej)}\int DM
e^{-iU(u,e)}e^{-iS_0(u)},
\ee
where the exponential over differential operator $\h{K}(ej)$, see
({\r{3.11}), gives perturbation series, functional $U(u,e)$, see
({\r{3.8}), describes interactions, the measure $DM$ is Diracian,
see (\r{i1}) and $S_0$ is the closed path action.

({\bf B}) If coordinate variable $u_c(\x,\eta)$ and corresponding
momentum $p_c(\x,\eta)$ obey the equations
\be
\{u_c,h_j\}=\f{\pa H_j}{\pa
p_c},~\{p_c,h_j\}=-\f{\pa H_j}{\pa u_c},~{\rm at}~j=0,
\l{w1}\ee
where $\{,\}$ is the Poisson bracket in the $(\x,\eta)$ space, if
\be
h_j(\x,\eta)=H_j(u_c,p_c),~h(\eta)\equiv h_0(\x,\eta),
\l{w2}\ee
where
\be
H_j(u,p)=\f{1}{2}p^2+v(u)-ju
\l{w3}\ee
is the total Hamiltonian, then (a) the transformed measure has the
form:
\be
DM(\x,\eta)=\d(E-h(T))\prod_t\d(\dot{\x}-\f{\pa
h_j}{\pa\eta}) \d(\dot{\eta}+\f{\pa h_j}{\pa\x}),
\l{w5}\ee
since $(u_c,p_c)$ are the solutions of incident (classical) Hamiltonian
equations, (b) the dimensions of vectors $(u_c,p_c)$ and of the
space $(\x,\eta)=W_G$ are arbitrary. This property is important
since the physical trajectory $u_c$ may occupy the space of dimension
${\rm dim}W_G\leq{\rm dim}T$, where $T$ is the incident phase
space. Moreover, (c) ${\rm dim}W_G$ may be even or odd.

({\bf C}) If the Green function $g(t-t')$ of equations
\be
\dot{\x}=\f{\pa h_j}{\pa\eta},~\dot{\eta}=-\f{\pa h_j}{\pa\x}
\l{w4}\ee
have the form:
\be
g(t-t')=\th(t-t'),~g(0)=1,
\l{gf}\ee
then the quantum corrections to semiclassical approximation are
accumulated on the boundaries of $W_G$:
\be
\R^q=\int_{\pa W_G}d\R^{q}.
\l{boun}\ee
This conclusion proves {\bf v.} The explicit form of $d\R^{q}$ will
be given below.

The generalization of formalism on the field theory, where
$u_c=u_c(\vec{x};\x,\eta)$, becomes evident noting (ii) and that the
space coordinate may be considered as the index (of special cell). By
the same reason (ii) and taking into account (iii) the formalism
allows to consider also the situation where
$(\x,\eta)=(\x,\eta)(x,t)$. Last one incorporates the gauge freedom.
So, in result, the mapping allows quantize the gauge theories without
Faddeev-Popov $ansatz$. This is important for non-Abelian gauge
theories, where the Faddeev-Popov ghosts and the Gribov's ambiguities
present the problem.

The field theories will be considered in subsequent publications. In
this paper I wish consider following questions.

-- Sec.2.  The differential measure for $\R(E)$ is derived.  The
connection between unitarity condition and d'Alembert's variational
principle is discussed. It is shown that our representation restores
the ordinary WKB perturbation series. The main general consequences
of functional Dirac measure are listed.

-- Sec.3. The transformations of the path-integral variables are
shown. The main purpose is to demonstrate the $(u,p)\rar(\x,\eta)$
canonical transformation. The coordinate transformations is
demonstrated also.

-- Sec.4. The main properties of new perturbation theory in the
invariant subspace (factor manifold) $W_G$ are shown considering
the simplest quantum-mechanical example. One can hope that this
properties stay useful for field-theoretical models also.

-- Sec.5. We solve the Coulomb problem to show explicitly the role
of the reduction for quantum systems as the consequence of ({\bf A,
B}) and ({\bf C}).

\section{Unitarity condition}

Purpose of this section is to show how the $S$-matrix unitarity
condition can be introduced into the path-integral formalism to find
measure (\r{i1}) $^{7}$.

The unitarity condition for the  $S$-matrix $SS^+ =I$ presents the
infinite set of nonlinear operator equalities:
\be iA A^* =A - A^*,
\l{2'}\ee
where $A$ is the amplitude, $S=I+iA$. Note, in this definition $A$ is
dimensionless. (Obviously the energy-momentum conservation
$\d$-functions are extracted from elements of $S$-matrix  and
then the net ones have the dimension of space $[x]$).  Expressing the
amplitude by the path integral one can see that the l.h.s. of
equality (\r{2'}) offers the double integral and, at the same time,
the r.h.s.  is linear combination of integrals.  Let us consider what
this linearization of product $AA^*$ gives.

Using the spectral representation of one-particle amplitude:
\be
A_1(u_1 ,u_2 ;E)=\sum_{n}\frac{\Psi^*_{n} (u_2)\Psi_n (u_1)}{E-E_n
+i\e}, ~~~\e \rar +0,
\l{3'}\ee
let us calculate
\be
\R(E)=\int du_1 du_2 A_1(u_1 ,u_2 ;E)A^*_1 (u_1 ,u_2 ;E).
\l{4'} \ee
The integration over end points $u_1$ and  $u_2$ is performed for
sake of simplicity only. Using ortho-normalizability of the wave
functions $\Psi_{n} (u)$ we find that
\be
\R(E)=\sum_{n}|\frac{1}{E-E_n+i\e}|^2 =\frac{\pi}{\e}\sum_{n}\d
(E-E_{n}).
\l{5'}\ee
Certainly, the last equality is nothing new but it is important to
note that $\R(E) \equiv 0$ for all $E\neq E_n$, i.e. that all
unnecessary contributions with $E\neq E_n$ were canceled by
difference in the r.h.s. of eq.(\r{2'}).  We will put just last
equality in (\r{5'}) in the basis of the approach.

We will build the perturbation theory for $\R(E)$ using the
path-integral definition of amplitudes $^{7}$. It leads to loss of
some information since the amplitudes can be restored in such
formulation with the phase accuracy only. Yet, that is quite enough
for calculation of the energy spectrum. So, instead of ${\rm
Sp}\{1/(E-H+i\e)\}$ $^{13}$, as follows from (\r{5'}), the
absorption parts $\sim \Im{\rm Sp}\{1/(E-H+i\e)\}$ would be
calculated only.

The statement that the unitarity condition unambiguously determines
the measure of path integral for $\R(E)$ looks like a tautology since
$\exp\{iS(u)\}$, where $S(u)$ is the action, is the unitary operator,
which shifts a system along the trajectory. (It is well known that
this unitary transformation is the analogy of tangent transformations
of classical mechanics $^{14}$.) I.e. the unitarity is already
included in the path integrals.

But the general path-integral solution contains unnecessary degrees
of freedom (unobservable states with $E\neq E_n$ in our example). Our
idea is to define the functional measure in such a way that the
condition of absence of unnecessary contributions be loaded from the
very beginning. Just in this case the unitarity becomes the
sufficient condition. Indeed, it will be shown that the equality
(\r{2'}) leads to $\d$-like functional measure, which unambiguously
determines the complete set of {\it classically permitted}
contributions.

Formal realization is simple: one should find, as it follows from
(\r{5'}), the linear path-integral representation for $\R(E)$.
Indeed, to see the integral form of our approach, let us use the
proper-time representation:
\be
A_1(u_1 ,u_2 ;E)=\sum_{n} \Psi_{n}
(u_1)\Psi^{*}_{n} (u_2)i
\int^{\infty}_{0}dTe^{i(E-E_{n}+i\e)T}
\l{6'} \ee
and insert it into (\ref{4'}):
\be
\R(E)=\sum_{n} \int^{\infty}_{0} dT_{+}dT_{-}
e^{-(T_{+}+T_{-})\e} e^{i(E-E_{n})(T_{+}-T_{-})}.
\l{7'} \ee

We  will introduce new time variables instead of $T_{\pm}$:
\be T_{\pm}=T\pm\tau,
\l{8'} \ee
where, as follows from Jacobian of transformation, $|\tau|\leq
T,~0\leq T\leq \infty$. But we can put $|\tau|\leq\infty$ since
$T\sim1/\e\rar\infty$ is essential in integral over
$T$. In result,
\be
\R(E)=2\pi\sum_{n}\int^{\infty}_{0} dT
e^{-2\e T} \int^{+\infty}_{-\infty}\f{d\tau}{\pi}
e^{2i(E-E_{n})\tau}.
\l{9'} \ee
In the last integral all contributions with $E\neq E_{n}$ are
canceled.  Note that the product of amplitudes $AA^*$ was
`linearized' after extraction of `virtual' time $\tau
=(T_{+}-T_{-})/2$. The physical meaning of such variables will be
discussed, see also $^{14}$. I.e. we would divide the dynamical
degrees of freedom on the `classical' (like $T=(T_{+}+T_{-})/2$) and
`quantum' (like $\tau$) ones. Such decomposition becomes possible if
the double integrals are considered.

\subsection{Dirac functional measure}

We will consider following path-integral:
\be
A_1(u_1 ,u_2;E)=i\int^{\infty}_{0}dT
e^{iET}\int_{u_1=u(0)}^{u_2=u(T)} Dx e^{iS_{C_+}(u)},
\l{3.1}\ee
where $C_+ $ is the Mills complex time contour $^{15}$:
\be
C_\pm:~t\rar t\pm i\e,~\e\rar+0,~0\leq t\leq T_\pm.
\l{mills}\ee
Calculating the probability to find a particle with energy $E$ ($\Im
E$ is not mentioned for sake of simplicity) we have:  \ba \R(E)=\int
du_1 du_2 |A|^2 =\int^{\infty}_{0} dT_+ dT_- e^{iE(T_+ -T_-)}
\t\n\\
\int^{u_+(T_+)=u_-(T_-)}_{u_+(0)=u_-(0)} D_{C_+}u_+
D_{C_-}u_-e^{iS_{C_+(T_+)}(u_+)-iS_{C_-(T_-)}(u_-)},
\l{3.2a}\ea
where, see (\r{mills}), $C_- (T)=C^{*}_{+}(T)$. Note that the total
action $S_\wp(u)\equiv(S_{C_+(T_+)}(u_+)-S_{C_-(T_-)}(u_-))$
describes the closed-path motion by definition.

New independent time variables $T$ and $\tau$ will be used again, see
(\r{8'}). The mean trajectory $u(t)=(u_+(t)+u_-(t))/2$ and the
deviation $e(t)$ from it will be introduced, $u_{\pm}(t)=u(t)\pm
e(t)$. Note, we assume that this linear transformations in the
path integrals may be performed.

We will consider $e(t)$ and $\tau$ as the fluctuating, virtual,
quantities and calculate the integrals over them perturbatively. In
the zero order over $e$ and $\tau$, i.e. in the semiclassical
approximation, $u$ is the classical path and $T$ is the total time of
classical motion.

The boundary conditions (see (\r{3.2a})) states the closed-path
motion. We would consider the boundary conditions for $e(t)$ only:
\be
e(0)=e(T)=0.
\l{3.6}\ee
Note the uniqueness of this solution if the integral over $\tau$ is
calculated perturbatively.

Extracting the linear over $e$ and $\tau$ terms from the closed-path
action $S_\wp$ and expanding over $e$ and $\tau$ the remainder terms:
\be
-\tilde {H}_T (u;\tau)=(S_{C_+ (T+\tau)}(u) - S_{C_- (T-\tau )}(u))+
2\tau H_T (u)-S_0(u),
\l{3.7}\ee
where $H_T(u)$ is the Hamiltonian at the time moment $T$, and
\be
-U_T (u,e)=(S_{C_+ (T)}(u+e)-S_{C_- (T)}(u-e))+2\Re\int_{C_+(T)}dt
(\ddot{u}+ v'(u))e
\l{3.8}\ee
we find that
\be
\R(E)=2\pi \int^{\infty}_{0}dTe^{-i\h{K}(\o ,\tau;j,e)} \int DM(u)
e^{-i\tilde{H}_T (u;\tau)-iU_T (u,e)+iS_0(u)}.
\l{3.10}\ee
Note the necessity of boundary condition (\r{3.6}) to find
(\r{3.10}). It allows to split the expansions over $\tau$ and $e$.

The  expansion over differential operators:
\be
\h{K}(\o ,\tau;j,e)=
\frac{1}{2}(\frac{\partial}{\partial\o}\frac{\partial}{\partial \tau}
+\Re\int_{C_+(T)}dt\frac{\delta}{\delta j(t)}\frac{\delta}{\d e(t)})
\l{3.11}\ee
will generate the perturbation series. We will assume that it exist at
least in Borel sense.

In (\r{3.10}) the functional measure
\be
DM(u)=\delta (E+\o -H_T(u))\prod_t du(t) \delta (\ddot{u}+v'(u)-j)
\l{3.12}\ee
unambiguously defines the complete set of contributions in the path
integral. The functional $\delta$-function is defined as follows:
\be
\prod_t
\delta (\ddot{u}+v'(u)-j)=(2\pi )^2 \int^{e(T)=0}_{e(0)=0}
\prod_{t}\frac{de(t)}{\pi}
e^{-2i\Re\int_{C_+}dt e(\ddot{x}+v'(u)-j)}
\l{*}\ee
$Note$, the phase in (\r{*}) stay real for arbitrary directions in
the complex plane of $e$. This explains why calculation of the modulo
square of amplitudes is important.

The physical meaning of this $\d$-function is following. We can
consider $(\ddot{u}+v'(u)-j)$ as the total force and $e(t)$ as the
virtual deviation from true trajectory $u(t)$. In classical mechanics
the virtual work must be equal to zero: $(\ddot{u}+v'(u)-j)e(t)=0$
(d'Alembert) $^{16}$ since the motion is time reversible. From this
evident dynamical principle one can find the `classical' equation of
motion:
\be
\ddot{u}+v'(u)=j,
\l{3.13}\ee
since $e(t)$ is arbitrary.

In quantum theories the virtual work usually is not equal to zero,
i.e.  the quantum motion is not time reversible since the quantum
corrections can shift the energy levels. But integration over $e(t)$,
with boundary conditions (\ref{3.6}), leads to the same result
(\r{3.13}).  So, in quantum theories the unitarity condition $^{7}$
play the same role as the d'Alembert's variational principle in
classical mechanics. We can conclude, the unitarity condition as the
dynamical principle establish the $time-local$ equilibrium between
classical (l.h.s. of (\r{3.13})) and quantum (r.h.s. of (\r{3.13}))
forces.

So, considering the double integral we may introduce integration over
two independent fields $u$ and $e$. Then, (i) integral over $e$ gives
the $\d$-function (\r{*}) and (ii) last one defines integral over
$u$. This definition of path integrals permits the Mills' analitical
continuation into complex time plane.

It should be underlined that the real-time field theory is
considered. We found actually that the real-time theories are simple,
see the functional measure (\r{i1}). This seems important since the
Wick rotation is practically noncontrollable if the symmetry is high
(symmetry content of a theory is sensitive to the space-time metrics
$^{17}$) and especially if the dynamical problems are solved.

\subsection{Comparison with WKB perturbation theory}

Let us consider now the representation (\r{3.10}). It is not hard to
show that it restores the perturbation theory of stationary phase
method.  For this purpose it is enough to consider the ordinary
integral:
\be
A(a,b)=\int^{+\infty}_{-\infty}\frac{dx}{(2\pi)^{1/2}}
e^{i(\frac{1}{2}ax^2+\frac{1}{3}bx^3)},
\l{3.14} \ee
with $\Im a \rar +0$ and $b>0$. Computing the `probability' $\R=|A|^2$
we find:
\be
\R(a,b)=e^{\frac{1}{2i}\hat{j}\hat{e}}\int^{+\infty}_{-\infty} dx
e^{-2(x^2 +e^2 )Im~a}e^{2i\frac{b}{3}e^3}\d (\Re a~x +bx^2+j).
\l{3.1''}\ee
The `hat' symbol means, as usual, the derivative over corresponding
quantity:  $\h{X}\equiv\pa/\pa X$. One should put the auxiliary
variables ($j,e$) equal to zero at the very end of calculations.

Performing the trivial transformation $e\rar ie$, $\hat{e}\rar
-i\hat{e}$ of auxiliary variable we find in the limit $\Im a=0$ that
the contribution of $x=0$ extremum (minimum) gives expression:
\be
\R(a,b)=\frac{1}{a}e^{-\frac{1}{2}\hat{j}\hat{e}}(1-4bj/a^2)^{-1/2}
e^{2\frac{b}{3}e^3}
\l{3.15'}\ee
and the expansion of operator exponent gives the asymptotic  series:
\be
\R(a,b)=\frac{1}{a}\sum^{\infty}_{n=0}(-1)^{n}\frac{(6n-1)!!}{n!}
(\frac{2b^4}{3a^6})^n,~~~(-1)!!=0!!=1.
\l{3.16'}\ee
This series is convergent in Borel sense.

$Note$, if $|A|$ only is interesting for us then
eq.(\r{3.1''}) may be considered as the definition of integral
(\r{3.14}).  By this reason one may put $\Im a =0$ from the very
beginning. We will consider this suggestion more carefully.

Let us calculate now $\R$ using the stationary phase method.
Contribution from the minimum $x=0$ gives $(\Im a=0)$:
$$
A(a,b)=e^{-i\hat{j}\hat{x}}e^{-\frac{i}{2a}j^2}e^{i\frac{b}{3}x^3}
(\frac{i}{a})^{1/2}.
$$
The corresponding `probability' is
\be
\R(a,b)=\frac{1}{a}e^{-\frac{1}{2}\hat{j}\hat{e}}e^{2\frac{b}{3}e^3}
e^{\frac{2b}{a^2}ej^2}
\l{3.17}\ee
This expression does not coincide with (\ref{3.15'}) but it leads to
the same asymptotic series (\r{3.16'}).

To find the representation (\r{3.17}) from (\r{3.15'}) the
transformation
\be
\d (\Re a~x +bx^2+j)=e^{-\f{i}{2}\h{j'}\h{e'}}e^{+2i(bx^2+j)e'}
\d (\Re a~x +j')
\l{trans}\ee
can be applied. Indeed, inserting this equality into (\r{3.15'}) we
find (\r{3.17}). The eq.(\r{trans}) is evident from
the Fourier transformation of $\d$-function.

$Note$, the transformation (\r{trans}) practically solves,
linearizing argument of $\d$-function, the problem of computation of
the determinant. This will be important considering functional
integrals. Moreover, it reflects the freedom in choice of terms in
which the perturbation theory in vicinity of nontrivial trajectories
in functional space is realized.

Just this property is the source of splitting:
\be
j(t)\rar (j_\th (t), j_h (t))
\l{i8}\ee
of the `Lagrange' source $j(t)$ onto set of sources to each
independent degree of freedom of the invariant subspace if the
transformation ({\bf A}) was performed. This splitting is
demonstrated in Appendix A. By this way the actually Hamiltonian
description is achieved in the invariant subspace.

\subsection{General properties of theory with $\d$-like measure}

The solution $x_j (t)$ of eq.(\r{3.13}) we would search expanding it
over $j(t)$:
$$
u_j (t)=u_c (t)+\int dt_1 G(t,t_1;u_c)j(t_1 )+...
$$
This is sufficient since $j(t)$ is  the auxiliary variable. In this
decomposition $u_c (t)$ is the strict solution of unperturbated
equation $\ddot{x}+v'(x)=0$ and $G(t,t';u_c)$ must obey eq.(\r{i3}).
Note that the functional $\d$-function in (\ref{*}) does not contain
the end-point values of time $t=0$ and $t=T$. This means that the
initial conditions to the eq.(\ref{3.13}) are not fixed and the
integration over them should be performed because of our definition
of $\R$.

The $\d$-likeness of measure allows to conclude that all strict
regular solutions (including trivial) of classical (unperturbated by
$j$) equation(s) of motion must be taken into account.

We must consider only `strict' solutions because of strict
cancellation of needless contributions. The $\d$-likeness of measure
means that the probability $\R(E)$ should contain a $sum$ over all
discussed solutions.  This is the main distinction of our unitary
method of quantization from stationary phase method: even having few
solution there is not interference terms in the sum over them in
$\R$. Note that the interference terms are absent independently from
solutions `nearness' in the functional space. This reflects the
orthogonality of Hilbert spaces builded on the various $u_c$ $^{1}$
and is the consequence of unitarity condition.

The solutions must be regular since the singular $u_c$ gives zero
contribution on $\d$-like measure.

It is evident that having the sum over contributions of various $u_c$
we must leave largest. This selection rule $^{7}$ is the
constructive definition of physical vacuum.

Summation over all solutions of classical equation of motion means
also necessity to take into account all topologically-equivalent
orbits $u_c$. This means integration over the volume $V_W$ of factor
space $W_G=G/\bar{G}$.  This naturally introduces integration over
zero-mode degrees of freedom.

So, our selection rule looks as follows: if there is not special
external constraints then in the sum over topologically nonequivalent
trajectories one should leave, up to the volume $V_W$, the
contribution defined in the highest factor manifold $G/\bar{G}$ if
$G$ is the theory symmetry group and $\bar{G}$ is the
$u_c$-invariance (sub)group of $G$ group.  (Note, $G$ may be wider
then the actions invariance group.) This selection rule means that
$\bar{G}$ should be the lowest (sub)group.

$Note$, the quantum corrections may violate our selection rule.

The Dirac measure defines the real-time motion only and is not
applicable for tunnelling processes since reflects the dynamical
equilibrium of `real' forces. The contributions from tunneling
processes should be added to the contributions defined by our
$\d$-like measure.  Then, following to our selection rule, we should
leave those contribution(s) which are proportional to the highest
volume $V_W$. So, our definition of measure is rightful if the
real-time contributions factor manifold have the highest dimension.
One can say in this case that the imaginary-time contributions are
realized on zero measure ($\sim 1/V_W$).

The explicit investigation of this condition is the nontrivial task
in spite of its seeming simplicity (the dimension of $G/\bar{G}$ is
defined by classical solution only). Actually we should know
$all$ classical orbits and the quantum corrections may `shrink' the
dimension of $W_G$. So, above selection rule gives the classification
only of mostly probable contributions. Following from our selection
rule we should start from the nontrivial solutions $u_c\neq0$ since
the volume of trivial $u_c=0$ is equal to zero.

\section{Canonical transformations}

It is evident that the measure (\r{3.12}) admits the canonical
transformations (the coordinate transformations are described in
Appendix B).  This follows from $\d$-likeness of measure. The phase
space differential measure has the form:
\be
DM(x,p)=\delta (E+\o -H_T (u))\prod_{t}du dp
\delta(\dot{u}-\frac{\partial H_j}{\partial p})
\delta(\dot{p}+\frac{\partial H_j}{\partial u}),
\l{3.18}\ee
where
\be
H_{j}=\frac{1}{2}p^2 +v(u)-ju
\l{3.19}\ee
is the time dependent through $j(t)$ total Hamiltonian.

The transformation may be performed inserting
\be
1=\int D\th Dh\prod_{t}\d (h-\frac{1}{2}p^2 -v(u))\d (\th -\int^{u}du
(2(h-v(u)))^{-1/2}).
\l{3.20} \ee
It is important that both differential measures in (\r{3.20}) and
(\r{3.18}) are $\delta$-like. This allows to change the order of
integration and firstly integrate over $(u,p)$. Calculating
result one can use the $\d$-functions of (\r{3.18}). In this case
the $\d$-functions of (\r{3.20}) will define the constraints. But if
we use the $\d$-functions of (\r{3.20}) the mapping $(u,p)\rar
(\th,h)$ is performed and the remaining $\d$-functions would define
motion in the factor space $W_G$.  We conclude that our transformation
takes into account the constraints since both ways must give the same
result. Note also, the transformation did not change the power of
manifolds since both measures, in $T$ and in $T^*G$, are $\d$-like.

We find by explicit calculations that:
\be
DM(\th ,h)=\d (E+\o-h(T))\prod_{t}d\th dh \d (\dot{\th}-\frac{\pa h_j}
{\pa h}) \d(\dot{h}+\f{\pa h_j}{\pa\th}),
\l{3.21}\ee
since considered transformation is canonical,
$\{h(u,p),\th (u,p)\}=1$, where
\be
h_j(\th,h) =h-ju_c (\th,h)
\l{3.22}\ee
is the transformed Hamiltonian and $u_c (\theta,h)$ is the classical
trajectory parametrized by $h$ and $\th$.

\subsection{General properties of the transformed perturbation
theory}

The transformed perturbation theory presents expansion over $1/g$ if
$u_c\sim 1/g$, where $g$ is the interaction constant.  Hence, we
construct the perturbation theory in the `strong coupling' limit. But
one should remind also that, generally, all solutions must be taken
into account.  This means that the perturbation theory for $\R(E)$
contains simultaneously both series over $g$ (from trivial solution
$u_c=0$) and over $1/g$, i.e.  the sum of week-coupling and
strong-coupling expansions. According to our selection rule we should
leave largest among them, i.e. start consideration from contributions
of nontrivial trajectories.

In the invariant subspace $W_G$ we must solve following equations of
motion:
\be
\dot{h}=j\f{\pa u_c(\th,h)}{\pa\th},~~~
\dot{\th}=1-j\f{\pa u_c(\th,h)}{\pa h}.
\l{3.23} \ee
They have a simple structure: the `propagator' in $W_G$ space is
simple $\Th$-function.

Indeed, expanding solutions of eqs.(\r{3.23}) over $j$, the zero
order solutions are $\th_0 =t_0 +t$ and $h_0 =const$. The first order
over $j$ gives:
$$
\dot{h}_1(t,t')=\d(t-t')\f{\pa u_c(\th_0,h_0)}{\pa\th_0},~~~
\dot{\th}_1(t,t')=-\d(t-t')\f{\pa u_c(\th_0,h_0)}{\pa h_0}.
$$
This leads to the first order equation for Green function $g(t-t')$:
\be
\dot{g}(t-t')=\d(t-t').
\l{3.24a}\ee
The solution of this equation introduces the time `irrevercibility':
\be
g(t-t')=\Th (t-t'),
\l{3.24}\ee
in opposite to causal particles propagator $G(t,t';u_c)$. But, as
will be seen below, the perturbation theory with Green function
(\r{3.24}) is time reversible.  Note also, that the solution
(\r{3.24}) is the unique and is the direct consequence of usual in
the quantum theories $i\e$-prescription.

The uncertainty is contained in the boundary value $g(0)$. We will
see that $g(0)=0$ excludes number of quantum corrections. By this
reason one should consider $g(0)\neq 0$. We will $assume$ that \be
g(0)=1
\l{3.24b}\ee
since this boundary condition to eq.(\r{3.24a}) is natural for local
theories.  We will use also following formal equalities:
\be
g(t-t')g(t'-t)=0,~~~1=g(t-t')+g(t'-t)
\l{3.24c}\ee
considering $g(t-t')$ as the distribution (generalized function).

$Note$, the important property (\r{boun}) of our perturbation theory
is the consequence of boundary condition (\r{3.24b}).

The property $\Im g(t)=0$ on the real time axis allows to conclude
that the perturbation theory in the $W_G$ space can be constructed on
the real-time axis.  This excludes natural for probabilistic
description doubling of degrees of freedom. But, for more confidence,
one should introduce the $i\e$-prescription and, extracting the
$\d$-function in the measure, to analyze the theory boundary
conditions in $\e=0$ limit. We will return to this question at the
end of this section.

\subsection{Splitting of Lagrange source $j(t)$}

Note now that $j\pa u_c/\pa \th$ and $j\pa u_c/\pa h$ in the r.h.s. of
(\r{3.23})  can be  considered as the new (renormalized) sources.
This allows to note that the mapping on the $W_G$ splits `Lagrange'
quantum force $j$ on a set of quantum forces individual to each
independent degree of freedom.

Indeed, the simple algebra gives (see Appendix A):
\ba
\R(E)=2\pi \int^{\infty}_{0}
dTe^{\frac{1}{2i}(\hat{\omega}\hat{\tau}+ \Re\int_{C_+} dt (\hat{j}_h
(t)\hat{e}_h (t)+ \hat{j}_{\th} (t)\hat{e}_{\th} (t)))}
\n \\ \times
\int Dh D\th e^{-i\tilde{H}(u_c ;\tau )-iV_T (u_c ,e_c )+iS_0(u_c)}
\n\\ \times
\delta (E+ \omega -h(T))\prod_{t} \delta (\dot{h} -j_h )\delta
(\dot{\th} -1 - j_{\th}),
\l{27b}\ea
where
\be
e_c =e_h \frac{\pa u_c}{\pa \th} -e_{\th} \frac{\pa u_c}{\pa h}
\equiv (e_h \h{\th}-e_\th\h{h})u_c.
\l{23b}\ee
$Note$, $e_c$ carry the symplectic structure of Hamiltonian equations
of motion, see (\r{3.21}), i.e.  $e_c$ is the invariant of canonical
transformations.

Hiding the $u_c (t)$ dependence in $e_c$ we solve the problem of
the functional determinants and simplify the equation of motion as
much as possible. Performing the shift:
$$
\th\rar\th+\th',~~~h\rar h+h',
$$
where
$$
\th'(t)=\int^{T}_{0}dt'g(t-t')j_\th(t'),~~
h'(t)=\int^{T}_{0}dt'g(t-t')j_h(t')
$$
we can consider $(\th',h')$ as the independent virtual variables:
\be
DM(h,\th )=\d (E+\o -h(T)-h'(T))\prod_{t}
dh(t) d\th(t) \d (\dot{h}(t)) \d (\dot{\theta}(t)-1)
\l{3.26}\ee
and new perturbations generating operator takes the form:
\be
\h{K}=\frac{1}{2}(\hat{\o}\hat{\tau}+ \int^T_0 dt_1 dt_2 \Theta (t_2
-t_1 ) (\hat{e}_h (t_1 )\hat{h}'(t_2 )+ \hat{e}_{\theta}(t_1
)\hat{\theta}' (t_2 )).
\l{3.27}\ee
In $U_T (u_c,e_c)$ we must change $h \rar (h + h')$ and $\th \rar
(\th +\th')$.

\subsection{Zero modes problem}

Noting that
\be
\int \prod_t dX(t)\d (\dot{X}(t))=\int dX(0)=\int dX_0
\l{zo}\ee
we see that the measure (\r{3.26}) coincide with the measure
of ordinary integrals over $h_0$ and $t_0$. Last one defines the
volume of translational mode. Note, using naively the WKB expansion
we should find $\R\sim V^2$, where $V$ is the zero modes volume,
since $\R\sim|A|^2$. But, as follows from (\r{3.26}) and (\r{zo}), we
may find only that $\R\sim V$. This evident discrepancy follows from
our rough analytical continuation on the real time axis: it may, as
was noted above, eliminate a doubling of degrees of freedom intrinsic
to considered approach.

Let us consider this question more carefully. Deriving explicit form
of the operator $\hat{\cal O}$ following boundary conditions was
applied:
\be
u_+(t\in\pa C_\pm)=u_-(t\in\pa C_\pm),
\l{q1}\ee
where $\pa C_\pm$ are boundaries on corresponding branches of the
total Mills time contour $C=C_++C_-$. Generally, performing canonical
mapping $(u,p)\rar(\th,h)$,
\be
(u,p)(t)|_{C_\pm}\rar(\th,h)(t)|_{C_\pm}
\l{q2}\ee
since one should hold the Mills contours memory. Then, noting that
the auxiliary variable $e(t\in\pa C_\pm)=0$, the boundary conditions
(\r{q1}) means following equalities:
\be
u_c(\th_+,h_+){(t\in\pa C_\pm)}=u_c(\th_-,h_-){(t\in\pa C_\pm)}
\l{q3}\ee
Inserting here the explicit value of $u_c$ we find the boundary
conditions for $(\th,h)|_\pm(t)$. Hence, the doubling of degrees of
freedom would disappeared iff (\r{q3}) leads to equality of
generalized coordinates $(\th,h)$ on the corresponding boundaries
$\pa C_\pm$. Contrary the doubling of degrees of freedom should be
taken into account.

We will find solving the equations (\r{3.26}) that the doubling of
degrees of freedom should be taken into account in definition of
initial data $(\th(0),h(0))$ only since the Green function of
transformed theory is nonsingular on the real time axis. So, if the
solution of (\r{q3}) gives, for instance, $h_+(0)=h_-(0)$ then the
doubling of scale degree of freedom $h(0)$ would disappear. Note
also, if the classical trajectory is periodic function then we may
choose the initial phases $\sim\th_\pm(0)$ independently.  Just this
effect takes into account the phase $S_0$ in (\r{3.7}).

From very beginning the measure $DM(u,p)$ is defined on the whole
Mills contour $C=C_++C_-$:
\be
DM(u,p)=\prod_{t\in C_+}\prod_{t\in C_-}...,
\l{q4}\ee
assuming corresponding generalization of $\d$-functions on the
complex arguments, see (\r{*}). This property should be conserved in
the transformed perturbation theory. Hence, if the boundary condition
(\r{q3}) will not lead to disappearance of the doubling, after
integrations we would have, instead of (\r{zo}), double integrals
$$
\int dX_+(0)dX_-(0).
$$

\section{Perturbation theory}

Let us consider motion in the action-angle phase space. Corresponding
perturbations generating operator has the form:
\be
\h{K}=\frac{1}{2}\int_0^T dtdt' \Theta (t-t') (\hat{I}(t)\hat{e}_I
(t')+\hat{\phi}(t)\hat{e}_{\phi} (t'))\equiv \h{K}_I +\h{K}_{\p}.
\l{3.32}\ee
The result of integration using last $\d$-function is
\be
\R(E)=2\pi \int^{\infty}_{0} dT  e^{-i\h{K}}\int^{2\pi}_{0}
\frac{d\phi_0}{\o (E)}e^{-iU_T (u_c ,e_c )},
\l{3.33}\ee
where
$$\o =\pa h(I_0)/\pa I_0$$
and $I_0=I_0(E)$ is defined by the algebraic equation:  $$E=h(I).$$
The classical trajectory
\be
u_c (t)=u_c (I_0(E)+I(t)-I(T), \phi_0
+\tilde{\o}t+\phi (t)),
\l{3.33'}\ee
where $$\tilde{\o}=\f{1}{t}\int^T_0 dt' \Th(t-t')\o (I_0 +I(t')).$$
The interaction `potential' $U_T$ depends from \be e_c=e_{\p}\f{\pa
u_c}{\pa I}- e_{I}\f{\pa u_c}{\pa \p}.  \l{3.33''}\ee

The operator (\ref{3.27}) contains unnecessary terms.  One can omit
the $\tau$ dependance since the closed-path motion is described. This
simplification was used in (\r{3.32}) and (\r{3.33}).

The operator $\h{K}$ is linear over $\h e_{\p}$, $\h e_{I}$. The
result of its action can be written in the form:
\be
\R(E)=2\pi\int^{\infty}_{0} dT \int^{2\pi}_{0} \frac{d\phi_0}{\o
(E)}:e^{-iU_T (u_c ,\h{e}_c/2i)}e^{iS_0(u_c)}:,
\l{a1}\ee
where
\be
\h{e}_c=\h{j}_{\p}\f{\pa u_c}{\pa I}- \h{j}_{I}\f{\pa u_c}{\pa \p}=
(\h{j}_\p\h{I}-\h{j}_I\h{\p})u_c=\int^T_0 dt' \th(t-t')
\{\h{\p}(t'),\h{I}(t)\}u_c(t)
\l{a2}\ee
since
\be
\h{j}_{X}(t)=\int^T
_0 dt' \Th (t'-t) \h{X}(t'),~~~X=\p ,I.
\l{a3}\ee
The colons in (\r{a1}) means 'normal product': the differential
operators must stay to the left of $all$ functions in expansion over
commutator
$$\{\h{\p}(t'),\h{I}(t)\}=\h{\p}(t')\h{I}(t)-\h{I}(t')\h{\p}(t).$$

Now we are ready to offer the important statement:
{\it If the eqs.}(\r{3.24}, \r{3.24b}, \r{3.24c}){\it are hold then
each term of perturbation theory in the invariant subspace $W_G$
can be represented as the sum of total derivatives over the subspace
$W_G$ coordinates.}

This statement directly follows from definition of perturbation
generating operator $\h{K}$ on the cotangent bundle (\r{3.27}) and of
homogeneity of the cotangent manifold in the classical
approximation. The proof of this statement is given in Appendix C.

So, we can conclude, contributions are defined by boundary values of
classical trajectory $u_c$ in the invariant subspace since the
integration over $X_0=(\x,\eta)_0$ is assumed, see (\r{3.33}), and
since contributions are the sum of total derivatives over $X_0$.

\section{H-atom problem}

Let us calculate now the integral:
\be
\R(E)=\int^{\infty}_0 dTe^{-i\h{K}(j,e)}\int DM(p,l,r,\vp)
e^{-iU_T(r,e)}e^{iS_0},
\l{21}\ee
where $\h{K}(j,e)$ was defined in (\r{3.29}) and $DM(p,l,r,\vp)$ in
(\r{3.28}).

Considering the Coulomb potential
\be
U_T(r,e)=\int^T_0
dt[\f{1}{((r+e_r)^2+r^2e_{\vp}^2)^{1/2}}-
\f{1}{((r-e_r)^2+r^2e_{\vp}^2)^{1/2}}+2\f{e_r}{r}]
\l{26k}\ee
describes the interaction.

We will restrict ourselves by the plane problem. Corresponding phase
space $T=(p,l,r,\vp)$ is 4-dimensional. But the classical flaw of this
problem can be parametrized by the angular momentum $l$,
corresponding angle $\vp$ and by the normalized on total Hamiltonian
Runge-Lentz vectors length $n$. So, we will demonstrate the mapping
($p$ is the radial momentum in the cylindrical coordinates):
\be
J_{l,n}:(p,l,r,\vp)\rar(l,n,\vp)
\l{i16}\ee
to construct the perturbation theory in the $W_C=(l,n,\vp)$ space.
I.e.  $W_C$ is not the symplectic space: $W_C=T^*G\times R^1$, where
$(l,\vp)\in T^*G$ is the symplectic space and $n\in R^1$.  Nevertheless
we start from the symplectic space adding to $n$ the auxiliary
canonical variable $\x$.

It is well known $^{18}$ that the consequence of hidden conservation
of the Runge-Lentz vector $\vec{N}$ is closeness of the Kepler orbits
independently from initial conditions.  In result the orbit is the
function of $|\vec{N}|$ only. The external field leads to
precession of $\vec{N}$ and the orbit should be parametrized by 4
parameters in this case. So, the reduction (\r{i16}) takes into
account the hidden symmetry of the Coulomb problem.

The bound state energies ($E<0$) in the Coulomb potential to
illustrate our idea will be calculated. This popular problem was
considered by many authors, using various methods $^{18}$. The
path-integral solution of this problem was offered in $^{19}$ using
the time-sliced method.

\subsection{General formalism of mapping}

We would consider now more general method of mapping in the $W_C$
space.  It is important to start from the  assumption that the
invariant subspace has symplectic structure of cotangent manifold
and its farther possible reduction to linear subspace $W_C$
(${\rm dim}(T^*G)\leq{\rm dim}(W_C)\leq{\rm dim}(T)$) would be
realized as the reduction of quantum degrees of freedom.

The first step of mapping consist in demonstration that the classical
trajectories belong to $T^*G$ completely. Let
\be
\D=\int \prod_t d^2\x d^2\eta \d (r-r_c(\x,\eta))\d (p-p_c(\x,\eta))
\d (l-l_c(\x,\eta))\d (\vp-\vp_c(\x,\eta))
\l{27k}\ee
be the functional of some functions $(r,p,\vp,l)(t)$ and $(\x,~\eta)$
are two-vectors.  Introducing this functional we realize the
transformation:
$$
(r,p,\vp,l)\rar(r,p,\vp,l)_c(\x,\eta),
$$
i.e. we want to `hide' the $t$ dependence into the four functions
$(\x,\eta)(t)$ introducing the composite functions
$(r,p,\vp,l)_c(\x,\eta)$. This four functions will be defined later.
The functions $(\x,\eta)(t)$ are arbitrary.

So, it is assumed that there exist such functions $(r,p,\vp,l)(t)$
for given $(r,p,\vp,l)_c$ that
\ba
\D_c=\int \prod_t d^2\bar{\x}
d^2\bar{\eta} \d(\f{\pa r_c}{\pa\x}\cdot\bar{\x}+\f{\pa r_c}{\pa\eta}
\cdot\bar{\eta})\d(\f{\pa p_c}{\pa\x}\cdot\bar{\x}+
\f{\pa p_c}{\pa\eta}\cdot\bar{\eta})
\t\n \\
\d(\f{\pa \vp_c}{\pa\x}\cdot\bar{\x}+\f{\pa \vp_c}{\pa\eta}
\cdot\bar{\eta})\d(\f{\pa l_c}{\pa\x}\cdot\bar{\x}+
\f{\pa l_c}{\pa\eta}\cdot\bar{\eta})\neq 0.
\l{28}\ea
Note that this is the condition for $(r,p,\vp,l)_c(\x,\eta)$ only.

To perform the mapping we insert
\be
1=\D/\D_c
\l{x}\ee
into (\r{21}) and integrate over $(r,p,\vp,l)(t)$. The proof of
equality (\r{x}) is following. It assumes that one always can find
$(\x,\eta)$ from four equalities
$(r,p,\vp,l)_c(\x,\eta)=(r,p,\vp,l)(t)$. Then, noting (\r{28}) and
using definition of $\d$-function, eq.(\r{x}) becomes evident.

In result of simple calculations (see Appendix D) we find that
\be
DM(\x, \eta)=\d(E-H_0)\prod_td^2\x d^2\eta
\d^2(\dot{\x}-\f{\pa h_j}{\pa\eta})
\d^2(\dot{\eta}+\f{\pa h_j}{\pa \x}),
\l{212'}\ee
where $H_0=H_0(\eta)$ is the classical Hamiltonian, $H_0=H_j$ at
$j=0$. It is the desired result of transformation of the measure for
given `generating' functions $(r,p,\vp,l)_c(\x,\eta)$. In this
case the `Hamiltonian' $h_j (\x,\eta)$ is defined by four equations
(\r{211}):
\ba
\{r_c,h_j\}-\f{\pa H_j}{\pa p_c}=0,~
\{p_c,h_j\}+\f{\pa H_j}{\pa r_c}=0,
\n \\
\{\vp_c,h_j\}-\f{\pa H_j}{\pa l_c}=0,~
\{l_c,h_j\}+\f{\pa H_j}{\pa \vp_c}=0.
\l{211x}\ea

But there is another possibility. Let us assume that
\be
h_j (\x, \eta)=H_j (r_c, p_c, \vp_c, l_c)
\l{213}\ee
and the functions $(r,p,\vp,l)_c(\x,\eta)$ are unknown. Then
eqs.(\r{211x}) are the equations for this functions. It is not hard to
see that the eqs.(\r{211x}) simultaneously with equations given by
$\d$-functions in (\r{212'}) are equivalent of incident equations if
the equality (\r{213}) is hold. So, incident dynamical problem was
divided on two parts. First one defines the trajectory in the $W_C$
space through eqs.(\r{211x}). Second one defines the dynamics, i.e.
the time dependence, through the equations in arguments of
$\d$-functions in the measure (\r{212'}).

We should consider $r_c,~ p_c,~ \vp_c,~ l_c$ as the
classical orbits in the $\x,~\eta$ parametrization. The desired
parametrization of them is well known (one can find it in
arbitrary textbook of classical mechanics):
\be
r_c=\f{\eta_1^2(\eta_1^2+\eta_2^2)^{1/2}}
{(\eta_1^2+\eta_2^2)^{1/2}+\eta_2\cos \x_1},~
p_c=\f{\eta_2\sin \x_1}{\eta_1(\eta_1^2+\eta_2^2)^{1/2}},~
\vp_c=\x_1,~l_c=\eta_1.
\l{214}\ee
At the same time,
\be
h_j=\f{1}{2(\eta_1^2+\eta_2^2)^{1/2}} -j_r r_c -j_\vp \vp_c
\equiv h (\eta)-j_r r_c -j_\vp \vp_x.
\l{215}\ee
Note that $\x_2$ is the irrelevant variable for classical flow
(\r{214}). This conclusion hides the assumption that the space is
flat and homogeneous, i.e. an external field may violate this
solution.

Our mapping contains two steps. We introduce the set of $ansatz$
functions $(r,p,\vp,l)_c$ assuming that the eqs.(\r{211x}) have
solution $h_j(\x,\eta)$ for arbitrary $j$ and that the condition
(\r{28}) is hold. For this purpose auxiliary variable $\x_2$ was
added assuming that $\pa r_c/\pa\x_2\sim\epsilon\rar0$. In result we
found the measure (\r{212'}) and
\be
\dot{\x}=\f{\pa h_j}{\pa\eta},~
\dot{\eta}=-\f{\pa h_j}{\pa \x}.
\l{treq}\ee
Having (\r{212'}) we may invert the problem assuming that just $h_j$
is known: $h_j=H_j+O(\epsilon)$, see (\r{213}). In this case the
eqs.(\r{211x}) gave $(r,p,\vp,l)_c$ and taking $\pa
r_c/\pa\x_2 \sim\epsilon=0$ this set is the classical flow.

\subsection{Reduction of quantum degrees of freedom}

Noting that the derivatives over $\x_2$ are equal to
zero we find that
\ba
DM(\x, \eta)=\d(E-h(T))\prod_td^2\x
d^2\eta \d(\dot{\x}_1-\o_1+j_r\f{r_c}{\pa\eta_1}) \t\n \\
\d(\dot{\x}_2-\o_2+j_r\f{r_c}{\pa\eta_2})
\d(\dot{\eta}_1-j_r\f{\pa r_c}{\pa \x_1} -j_\vp)
\d(\dot{\eta}_2),
\l{217}\ea
where
\be
\o_i=\f{\pa h(\eta)}{\pa\eta_i}
\l{218}\ee
are the conserved in classical limit $j_r=j_\vp =0$ velocities in
the $W_C$ space.

It is seen from (\r{217}) that the length of Runge-Lentz vector is not
perturbated by the quantum forces $j_r$ and $j_{\vp}$. To investigate
the consequence of this fact it is necessary to project this forces on
the axis of $W_C$ space. This means splitting of $j_r,~j_{\vp}$ on
$j_\x,~j_\eta$. Then noting that the last $\d$-function in (\r{217})
is source-free, we find the same representation as (\r{21}) but with
\be
\h{K}(j,e)=\int^T_0 dt (\h{j}_{\x_1}\h{e}_{\x_1}+
\h{j}_{\x_2}\h{e}_{\x_2}+ \h{j}_{\eta_1}\h{e}_{\eta_1}), \l{31}\ee
where the operators $\h{j}$ are defined by the equality:
\be
\h{j}_X (t)=\int^T_0 dt' \theta(t- t')\h{X}(t')
\l{44}\ee
and $\theta(t- t')$ is the Green function of our perturbation theory.

We should change also
\be
e_r\rar e_c=e_{\eta_1}\f{\pa r_c}{\pa \x_1}-
e_{\x_1}\f{\pa r_c}{\pa \eta_1}-
e_{\x_2}\f{\pa r_c}{\pa \eta_2},~~e_\vp\rar e_{\x_1}
\l{32}\ee
in the eq.(\r{26k}). The differential measure takes the simplest form:
\ba
DM(\x, \eta)=\d(E-h(T))\prod_td^2\x d^2\eta
\d(\dot{\x}_1-\o_1-j_{\x_1})
\d(\dot{\x}_2-\o_2-j_{\x_2})
\n \\\times
\d(\dot{\eta}_1-j_{\eta_1})
\d(\dot{\eta}_2).
\l{33}\ea

Note now that the $\x, \eta$ variables are contained in $r_c$ only:
$r_c= r_c (\x_1, \eta_1, \eta_2)$. Then the action of the operator
$\h{j}_{\x_2}$ gives identical to zero contributions into
perturbation theory series. And, since $\h{e}_{\x_2}$ and
$\h{j}_{\x_2}$ are conjugate operators, see (\r{31}), we must put
$j_{\x_2}=e_{\x_2}=0$. This conclusion ends the reduction:
\be
\h{K}(j,e)=\int^T_0 dt (\h{j}_{\x_1}\h{e}_{\x_1}+
\h{j}_{\eta_1}\h{e}_{\eta_1}),
\l{34}\ee
\be
e_c=e_{\eta_1}\f{\pa r_c}{\pa \x_1}-e_{\x_1}\f{\pa r_c}{\pa \eta_1}.
\l{34'}\ee
Using (\r{zo}) the measure takes the form:
\be
DM(\x, \eta)=\d(E-h(T))d\x_2 d\eta_2\prod_td\x_1 d\eta_1
\d(\dot{\x}_1-\o_1-j_{\x_1})
\d(\dot{\eta}_1-j_{\eta_1})
\l{35}\ee
since $r_c$ is $\x_2$ independent.

\subsection{Topological analyses}

One can see from (\r{35}) that the reduction can not solve the H-atom
problem completely: there are nontrivial corrections to the orbital
degrees of freedom $(\x_1,\eta_1)$. By this reason we should consider
the expansion over $\h{K}$.

Using last $\d$-functions in (\r{35}) we find, see also $^{12}$
(normalizing $\R(E)$ on the integral over $\x_2$):
\be
\R(E)=\int^\infty_0 dT e^{-i\h{K}(j,e)}\int dM e^{iS_0-iU_T(r_c,e)},
\l{41}\ee
where
\be
dM=\f{d\x_1 d\eta_1}{\o_2(E)}.
\l{42}\ee
The operator $\h{K}(j,e)$ was defined in (\r{34}) and
\be
U_T(r_c,e_c)=+\int^T_0 dt[\f{1}{((r_c+e_c)^2+
r_c^2e_{\x_1}^2)^{1/2}}-
-\f{1}{((r_c-e_c)^2+r_c^2e_{\x_1}^2)^{1/2}}+2\f{e_c}{r_c}]
\l{a}\ee
with $e_c$ defined in (\r{34'}) and
\be
r_c(t)=r_c(\eta_1 +\eta(t), \bar{\eta}_2(E,T), \x_1+\o_1(t)
+\x(t)),
\l{43}\ee
where $\bar{\eta}_2(E,T)$ is the solution of equation $E=h(\eta)$.

The integration range over $\x_1$ and $\eta_1$ is as follows:
\be
\pa W_C: 0\leq \x_1 \leq 2\pi,~~-\infty \leq \eta_1 \leq +\infty.
\l{45}\ee
First inequality defines the principal domain of the angular variable
$\vp$ and second ones take into account the clockwise and
anticlockwise motions of particle on the Kepler orbits, $|\eta_1|=
\infty$ is the bifurcation line. Note, this excludes the singularity
at $r=0$.

We can write:
\be
\R(E)=\int^\infty_0 dT \int dM :e^{-iU_T(r_c,\h{e})}e^{iS_0}:
\l{47}\ee
since the operator $\h{K}$ is linear over $\h{e}_{\x_1},
\h{e}_{\eta_1}$.  The colons means our `normal product' and
$U_T(r_c,\h{e})$ is the functional of operators:
\be
2i\h{e}_c=\h{j}_{\eta_1}\f{\pa r_c}{\pa \x_1}-
\h{j}_{\x_1}\f{\pa r_c}{\pa \eta_1},~~2i\h{e}_{\x_1}=\h{j}_{\x_1}.
\l{48}\ee
Expanding $U_T(r_c, \h{e})$ over $\h{e}_c$ and $\h{e}_{\eta_1}$ we
find:
\be
U_T(r_c,\h{e})=2\sum_{n+m \geq 1}C_{n,m}\int^T_0 dt
\f{\h{e}_c^{2n+1}\h{e}_{\eta_1}^m}{r_c^{2n+2}},
\l{46}\ee
where $C_{n,m}$ are the numerical coefficients. We see that the
interaction part presents expansion over $1/r_c$ and, therefore, the
expansion over $U_T$ generates an expansion over $1/r_c$.

In result,
\be
\R(E)=\int^\infty_0 dT \int dM \{e^{iS_0 (r_c)} +
B_{\x_1}(\x_1, \eta_1) +
B_{\eta_1}(\x_1, \eta_1)\}.
\l{49}\ee
The first term is the pure semiclassical contribution and last ones
are the quantum corrections. They can be written as the total
derivatives:
\be
B_{\x_1}(\x_1, \eta_1)=\f{\pa}{\pa
\x_1}b_{\x_1}(\x_1, \eta_1),~~ B_{\eta_1}(\x_1, \eta_1)=\f{\pa}{\pa
\eta_1}b_{\eta_1}(\x_1, \eta_1).
\l{410}\ee
This means that the mean value of quantum corrections in the $\x_1$
direction are equal to zero:
\be
\int^{2\pi}_0 d\x_1 \f{\pa}{\pa \x_1}b_{\x_1}(\x_1,
\eta_1) =0
\l{411}\ee
since $r_c$ is the closed trajectory independently from initial
conditions.

In the $\eta_1$ direction the motion is classical:
\be
\int^{+\infty}_{-\infty} d\eta_1 \f{\pa}{\pa \eta_1}
b_{\eta_1}(\x_1, \eta_1)=0
\l{412}\ee
since (i) $b_{\eta_1}$ is the series over $1/r_c^2$ and (ii) $r_c
\rar \infty$ when $|\eta_1| \rar \infty$. Therefore,
\be
\R(E)=\int^\infty_0 dT \int dM e^{iS_0 (r_c)}.
\l{413}\ee
This is the desired result.

Noting that
$$
S_0 (r_c)= kS_1 (E),~~k=0, \pm 1, \pm 2,...,
$$
where $S_1 (E)$ is the action over one classical period $T_1$:
$$
\frac{\partial S_1 (E)}{\partial E}=T_1 (E),
$$
and using the identity $^{7}$:
$$
\sum^{+\infty}_{-\infty} e^{inS_1 (E)} =
2\pi \sum^{+\infty}_{-\infty}\d (S_1 (E) - 2\pi n),
$$
we find, normalizing on zero-modes volume, that
\be
\R(E)=\pi \sum_{n} \d (E + 1/2n^2).
\l{416}\ee

\section{Conclusion}

Described approach is based on three `whales'. They are (i) the
definition of observables in quantum theories as the modulo square of
amplitudes, (ii) the description of quantum processes as the
transformation induced by unitary operator $\exp\{iS(x)\}$, where
$S(x)$ is the classical action and (iii) the unitarity condition as
the principle which determines connection between quantum dynamics
and classical measurement (optical theorem). Less principal
assumptions, usually taken `by treaty', that the quantum perturbations
are switched on adiabatically, and the Feynman's $i\e$-prescription,
were used also.

The formalism in terms of observables only was considered to use all
above fundamental principles. It must be noted that we are forced to
work in terms of observables $\R(E)$ since, this was mentioned above,
the transformation mix the degrees of freedom in such a way that it
is impossible return to the habitual amplitudes formalism, writing
$\R\sim |A|^2$.

Offered approach should be considered as the useful technical trick
(probably not unique) helping to calculate the observables if the
complicated topologies should be taken into account and the
corresponding vacuum is so complicated that its quantitative
description is a hopeless task.

\vspace{0.4in}
{\Large \bf Acknowledgement}

I would like to thank my colleagues in the Institute of Physics
(Tbilisi), Institute of Mathematics (Tbilisi) and especially
A.Ushveridze and I.Paziashvili for helpful discussions. The
help of L.Lipatov and E.Levin in realization of the approach was
crucial. The paper was compiled in the JINR (Dubna, Russia).  I
would like to thank V.G.Kadyshevsi for fruitful interest to described
technique and underling idea. Further extension on the field theory
and possible applications in the particles physics was done with
A.Sissakian (see hep-th/9811160) and will be published. The work was
partly granted by Georgian Acad. of Sciences.

\appendix\section{Source cotangent foliation}\0
\renewcommand{\theequation}{A.\arabic{equation}}

Let us consider the perturbation-generating operators $\h K$ action
to show the splitting mechanism of the source $j(t)$:
\ba
e^{-i\f{1}{2}\Re\int_{C_+}
dt \hat{j}(t)\hat{e}(t) }e^{-iU_T (u_c,e)} \prod_{t} \d (\dot{h}
-j\frac{\pa u_c}{\pa \th}) \d (\dot{\th} -1 +j\frac{\pa u_c}{\pa h})=
\n \\ \int De_h De_{\th}e^{2i\Re\int_{C_+}dt (e_h
\dot{h}+e_{\th}(\dot{\th}-1))} e^{-iU_T (u_c,e_c )},
\l{22}\ea
where
\be e_c =e_h \frac{\pa u_c}{\pa \th} -e_{\th} \frac{\pa u_c}{\pa h}
\equiv (e_h \h{\th}-e_\th\h{h})u_c.  \l{23}\ee
The integrals  over
$(e_h ,e_{\th})$ will be calculated perturbatively:
\ba
e^{-iU_T
(u_c,e_c )}=\sum^{\infty}_{n_h ,n_{\th} =0}\frac{1}{n_h !n_{\th} !}
\int \prod^{n_h }_{k=1}(dt_k e_h (t_k))\prod^{n_{\th} }_{k=1}(dt'_k
e_{\th} (t'_k)) \t\n \\P_{n_h ,n_{\th}} (u_c ,t_1
,...,t_{n_h},t'_1,...,t_{n_{\th}}),
\l{24} \ea
where \be P_{n_h
,n_{\th}} (u_c ,t_1 ,...,t_{n_h},t'_1,...,t_{n_{\th}})= \prod^{n_h
}_{k=1} \hat{e}'_h (t_k) \prod^{n_{\th} }_{k=1} \hat{e}'_{\th}
(t'_k)e^{-iU_T (u_c,e'_c )} \l{25} \ee with $e'_c \equiv e_c (e'_h
,e'_{\th} )$ and the derivatives in this equality are calculated at
$e'_h =0$, $e'_{\th}=0$. At the same time, \ba \prod^{n_h }_{k=1} e_h
(t_k)\prod^{n_{\th} }_{k=1} e_{\th} (t'_k)= \prod^{n_h }_{k=1}
(i\hat{j}_h (t_k))\prod^{n_{\th} }_{k=1} (i\hat{j}_{\th} (t'_k)) \n
\\ \times e^{-2i\Re\int_{C_+} dt (j_h (t)e_h
(t)+j_{\th}(t)e_{\th}(t))}.
\l{26}\ea
The limit $(j_h , j_{\th})=0$
is assumed. Inserting (\ref{25}), (\ref{26}) into (\ref{22}) we find
new representation for $\R(E)$:
\ba \R(E)=2\pi \int^{\infty}_{0}
dTe^{iS_0(u_c)}e^{\frac{1}{2i}(\hat{\omega}\hat{\tau}+ \Re\int_{C_+}
dt (\hat{j}_h (t)\hat{e}_h (t)+ \hat{j}_{\th} (t)\hat{e}_{\th} (t)))}
\t\n \\ \int Dh D\th e^{-i\tilde{H}(u_c ;\tau )-iU_T (u_c ,e_c
)} \t\n\\ \delta (E+ \omega -h(T))\prod_{t} \delta (\dot{h} -j_h
)\delta (\dot{\th} -1 - j_{\th}) \l{27}\ea in which the `energy' and
the `time' quantum degrees of freedom are splitting.

\section{Coordinate transformations}\0
\renewcommand{\theequation}{B.\arabic{equation}}

Let us consider the coordinate transformations. For instance, the
two dimensional model with potential
$v=v((x^{2}_{1}+x^{2}_{2})^{1/2})$ is simplified considering it in
the cylindrical coordinates $x_1 =r \cos\phi$, $x_2 =r \sin\phi$.
Note, this transformation is not canonical.

Starting from flat space with trivial metric tensor $g_{\mu\nu}$ and
inserting
\be
1=\int Dr D\p \prod_t \d
(r-\sqrt{x_1^2 +x_2^2}) \d (\p - \arctan \f{x_2}{x_1})
\l{coord}\ee
we find the measure in the cylindrical coordinates:
\ba
D^{(2)}M(r,\p )=\d (E+\o -H_T (r,\p))\times
\n\\
\prod_t dr d\p r^2 (t) \d (\ddot{r}-\dot{\p}^2 r+v'(r)-j_r)\d (\pa_t
(\dot{\p}r^2 )-rj_{\p}),
\l{3.28}\ea
where $v'(r)=\pa v(r)/\pa r$ and $j_r$, $j_{\p}$ are the components
of $\vec{j}$ in the cylindrical coordinates.

The perturbation generating operator has the form:
\be
\h{K}(j,e)=\frac{1}{2}\{\hat{\o}\hat{\tau}+Re\int_{C_+} dt
(\hat{j}_{r}(t)\hat{e}_{r}(t)+\hat{j}_{\phi}(t)\hat{e}_{\phi}(t))\}
\l{3.29}\ee
and in $U_T (x,{e})$ we must change ${e}$ on ${e}_c$
with components
\be
e_{c,1} =e_r \cos\p -re_{\p}\sin\p,~~~
e_{c,2}=e_r \sin\p +re_{\p} \cos\p.
\l{3.30}\ee
Note, $e_\p$ was arise in product with $r$.

The transformation looks quite classically but the measure (\r{3.28})
and perturbation generating operator (\r{3.29}) can not be derived by
$naive$ coordinate transformation of initial path integral for
amplitude. This becomes evident noting that transformed
representation for $\R(E)$ can not be written in the product form
$\sim AA^*$ of two functional integrals.

It is interesting also to find the measure starting from the curved
space with the Lagrangian
\be
L=\f{1}{2}g_{\mu\nu}(y)\dot{y}^\mu\dot{y}^\nu -{v}(y)
\l{b6'}\ee
It is enough to consider the kinetic term only since, to find the
Dirac measure, we should extract the odd over $e$ terms from the
`closed-path' action $S_T (y+e)-S_T (y-e)$. This procedure is
`trivial' for potential term. The lowest over $e_\mu$ part of the
kinetic term have the form:
\be
2\{g_{\la\mu}\ddot{y}^\mu +\Ga_{\la,\mu\nu}\dot{y}^\mu\dot{y}^\nu\}.
\l{b7}\ee
Therefore, the semiclassical approximation is restored.

To find the quantum corrections we should linearize at least the
$O(e^3)$ term in the exponent
$$
\exp\{\Re\int dt g_{\la,\mu\nu} e^\la
\dot{e}^\mu\dot{e}^\nu\}.
$$
This is possible noting that
$$
e^\mu(t')\h{e}'_\mu(t')\dot{e}'^\nu(t)=e^\mu(t')\d_{\mu\nu}
\pa_{t'}\d(t-t')=\dot{e}^\nu \d(t-t').
$$
In result,
\be
DM(y)=\sqrt{|g(y+e)||g(y-e)|}\prod_\la\prod_t dy_\la \d
(g_{\la\mu}\ddot{y}^\mu +\Ga_{\la,\mu\nu}\dot{y}^\mu\dot{y}^\nu
+v_\la (y) -j_\la).
\l{b8}\ee
where $v_\la (y)=\pa_\la v(y)$ and $\Ga_{\la,\mu\nu}$ is the
Christoffel index.  The perturbations generating operator $\h{K}$ and
the weight functional $U_T(y;e)$ have the standard form.

\renewcommand{\theequation}{C.\arabic{equation}}
\section{Extraction of total derivatives}\0

By definition $U_T$ is the odd over $\h{e}_c$ local functional:
\be
U_T (u_c ,\h{e}_c )=2\int^T _0 \sum^{\infty}_{n=1} (\h{e}_c
(t)/2i)^{2n+1} v_n (u_c),
\l{a4}\ee
where $v_n (u_c)$ is some $function$ of $u_c$. Inserting (\r{a2}) we
find:
\be
:e^{-iU_T (u_c ,\h{e}_c
)}:=\prod^{\infty}_{n=1}\prod^{2n+1}_{k=0}
:e^{-iU_{k,n}(\h{j},u_c)}:, \l{a5}\ee where \be
U_{k,n}(\h{j},u_c)=\int^{T}_{0}dt (\h{j}_{\p}(t))^{2n-k+1}
(\h{j}_{I}(t))^{k}b_{k,n}(u_c).
\l{a6}\ee
Explicit form of the function $b_{k,n}(u_c)$ is not important.

Using definition (\r{a3}) it easy to find:
$$ \h{j}(t_1)b_{k,n}(u_c
(t_2))=\Th (t_1 -t_2) \pa b_{k,n}(u_c)/\pa X_0 $$
since $u_c =u_c (X(t)+X_0)$, see (\r{3.33'}), or
\be
\h{j}_{X,1}b_2=\Th_{12}\pa_{X_0}b_2
\l{a7}\ee
since indices $(k,n)$ are not important.

Let as start consideration from the first term with $k=0$. Then
expanding $\h{U}_{0,n}$ we describe the angular quantum fluctuations
only. Noting that $\pa_{X_0}$ and $\h{j}$ commute we can consider
lowest orders over $\h{j}$. The typical term of this expansion is
(omitting index $\p$) \be \h{j}_1 \h{j}_2 \cdots \h{j}_m b_1 b_2
\cdots b_m.  \l{a8}\ee It is enough to show that this quantity is the
total derivative over $\p_0$.  The number $m$ counts an order of
perturbation, i.e. in $m$-th order we have $(\h{U}_{0,n})^m$.

$m=1$. In this approximation we have, see (\r{a7}), \be \h{j}_1 b_1 =
\Th_{11}\pa_0 b_1 =\pa_0 b_1 \neq 0.  \l{a9}\ee Here the definition
(\r{3.24a}) was used.

$m=2$. This order is less trivial:
\be
\h{j}_1 \h{j}_2 b_1 b_2
=\Th_{21} b^2_1 b_2 + b^1_1 b^1_2 +\Th_{12}b_1 b^2_2, \l{a10}\ee
where \be b^n_i \equiv \pa^n b_i.  \l{a11}\ee Deriving (\r{a10}) the
first equality in (\r{3.24c}) was used. At first glance (\r{a10}) is
not the total derivative. But inserting $$ 1=\Th_{12}+\Th_{21},$$
(see the second equality in (\r{3.24c})) we can symmetrize it:  \ba
\h{j}_1 \h{j}_2 b_1 b_2 =\Th_{21}( b^2_1 b_2 + b^1_1 b^1_2)+
\Th_{12}(b_1 b^2_2+b^1_1 b^1_2)= \n \\ \pa_0 (\Th_{21} b^1_1 b_2
+\Th_{12}b_1 b^1_2) \l{a12}\ea since the explicit form of function
$b$ is not important. So, the second order term can be reduced to the
total derivative also. Note, that the contribution (\r{a12}) contains
the sum of all permutations. This shows the `time reversibility' of
the constructed perturbation theory.

Let us consider now expansion over $\h{U}_{k,m}$, $k\neq 0$. The
typical term in this case is \be \h{j}^1_1 \h{j}^1_2 \cdots \h{j}^1_l
\h{j}^2_{l+1} \h{j}^2_{l+2} \cdots \h{j}^2_m b_1 b_2 \cdots b_m ,~~~
0<l<m, \l{a16}\ee where, for instance, $$ \h{j}^1_k \equiv \h{j}_I
(t_k),~~~\h{j}^2_k \equiv \h{j}_{\p} (t_k)$$ and \be \h{j}^i_1 b_2
=\Th_{12}\pa^i_0 b_2 \l{a17}\ee instead of (\r{a7}).

$m=2$, $l=1$. We have in this case:  \ba \h{j}^1_1 \h{j}^2_2 b_1 b_2
= \Th_{21}(b_2 \pa^1_0 \pa^2_0 b_1 + (\pa^2_0 b_2)(\pa^1_0 \pa^2_0
b_1))+ \n\\\Th_{12}(b_1 \pa^1_0 \pa^2_0 b_2 + (\pa^2_0 b_2)(\pa^1_0
\pa^2_0 b_1)) =\pa^1_0 (\Th_{21} b_2 \pa^2_0 b_1 + \Th_{12}b_1
\pa^2_0 b_2)+\n\\ \pa^2_0 (\Th_{21} b_2 \pa^1_0 b_1 + \Th_{12}b_1
\pa^1_0 b_2).  \l{a18}\ea Therefore, we have the total-derivative
structure yet.

This important property of new perturbation theory is conserved in
arbitrary order over $m$ and $l$ since the time-ordered structure
does not depend from upper index of $\h{j}$, see (\r{a17}).

\renewcommand{\theequation}{D.\arabic{equation}}
\section{General formalism of mapping}\0

The resulting measure looks as follows:
\ba
DM(\x, \eta)=\f{1}{\D_c}\d (E-H_0)\prod_t d^2\x d^2\eta
\d (\dot{r_c}-\f{\pa H_j}{\pa p_c})\times
\t\n \\
\d (\dot{p_c}+\f{\pa H_j}{\pa r_c})
\d (\dot{\vp_c}-\f{\pa H_j}{\pa l_c})
\d (\dot{l_c}+\f{\pa H_j}{\pa \vp_c}),
\l{29}\ea
Note that the parametrization $(r_c,p_c,\vp_c,l_c)(\x,\eta)$ was not
specified.

A simple algebra gives:
\ba
DM(\x, \eta)=\f{\d(E-H_0)}{\D_c}\prod_td^2\x d^2\eta
\int \prod_t d^2\bar{\x}d^2\bar{\eta}
\t\n \\
\d^2(\bar{\x}-(\dot{\x}-\f{\pa h_j}{\pa\eta}))
\d^2(\bar{\eta}-(\dot{\eta}+\f{\pa h_j}{\pa \x}))
\t\n \\
\d(\f{\pa r_c}{\pa\x}\cdot\bar{\x}+\f{\pa r_c}{\pa\eta}
\cdot\bar{\eta} +
\{r_c,h_j\}-\f{\pa H_j}{\pa p_c})
\t\n\\
\d(\f{\pa p_c}{\pa\x}\cdot\bar{\x}+\f{\pa p_c}{\pa\eta}
\cdot\bar{\eta} +
\{p_c,h_j\}+\f{\pa H_j}{\pa r_c})
\t\n \\
\d(\f{\pa \vp_c}{\pa\x}\cdot\bar{\x}+\f{\pa \vp_c}{\pa\eta}
\cdot\bar{\eta}+
\{\vp_c,h_j\}-\f{\pa H_j}{\pa l_c})
\t\n \\
\d(\f{\pa l_c}{\pa\x}\cdot\bar{\x}+\f{\pa l_c}{\pa\eta}
\cdot\bar{\eta}+
\{l_c,h_j\}+\f{\pa H_j}{\pa \vp_c}).
\l{210}\ea
The Poisson notation:
$$
\{X,h_j\}=\f{\pa X}{\pa \x}\f{\pa h_j}{\pa \eta}-
\f{\pa X}{\pa \eta}\f{\pa h_j}{\pa \x}
$$
was introduced in (\r{210}).

We will define the `auxiliary' quantity $h_j$ by following
equalities:
\ba
\{r_c,h_j\}-\f{\pa H_j}{\pa p_c}=0,~
\{p_c,h_j\}+\f{\pa H_j}{\pa r_c}=0,
\n \\
\{\vp_c,h_j\}-\f{\pa H_j}{\pa l_c}=0,~
\{l_c,h_j\}+\f{\pa H_j}{\pa \vp_c}=0.
\l{211}\ea
Then the functional determinant $\D_c$ is canceled and
\be
DM(\x, \eta)=\d(E-H_0(\eta))\prod_td^2\x d^2\eta
\d^2(\dot{\x}-\f{\pa h_j}{\pa\eta})
\d^2(\dot{\eta}+\f{\pa h_j}{\pa \x}),
\l{212}\ee

\end{document}